# Crystallization Behavior of ZBLAN Glass Under Combined Thermal and Vibrational Effects: Part I - Experimental Investigation

Ayush Subedi[1], Anthony Torres[1], Jeff Ganley[2], Ujjwal Dhakal[3]

[1]Materials, Science, Engineering, and Commercialization (MSEC), Texas State University, San Marcos, Tx 78666, United States

[2]Air Force Research Labs, Space Vehicles Directorate

[3]Department of Physics, Texas State University, San Marcos, Tx 78666, United States

**Abstract**

ZBLAN ($ZrF_4$-$BaF_2$-$LaF_3$-$AlF_3$-$NaF$) glass is a promising material for infrared optical fibers due to its wide transmission window and low theoretical attenuation. However, its strong tendency to crystallize during processing limits optical performance. While microgravity suppresses crystallization, the influence of mechanical vibration under terrestrial conditions remains poorly understood. This study systematically investigates the effect of vibration on the crystallization behavior of ZBLAN under controlled thermal environments. Using a custom-built heating and vibration apparatus, ZBLAN samples were subjected to varied temperatures and vibration levels, and their crystallization onset and morphological evolution were examined through optical microscopy, SEM, EDS, and AFM analyses. Results show that vibration reduces the crystallization onset temperature, indicating enhanced atomic mobility and nucleation kinetics. Progressive morphological transitions from needle-like to bow-tie and feather-like crystals were observed with increasing temperature and vibration intensity. Surface roughness measurements corroborate these findings, revealing a significant increase in nanoscale roughness in crystallized regions. Although short exposure duration and intermittent thermal decoupling introduced some variability, the overall results confirm that vibration acts as a direct facilitator of nucleation rather than a thermally mediated effect. This work provides new insight into vibration-induced crystallization mechanisms in fluoride glasses and establishes the experimental foundation for subsequent modeling and apparatus optimization studies under both terrestrial and microgravity conditions.

**Introduction**

In the last few decades, the glass fiber industry has had a substantial increase in interest due to the growing demand for high-speed telecommunication. According to the New York Times, the average person consumes approximately 34 Gigabytes of information per day [1]. This is fueled by high-speed fiber-based internet connections that deliver on demand content to consumers through social media, streaming services, online shopping, etc.  This is all made possible through glass (optical) fiber, most referred as a "fiber" internet connection. Glass fiber has emerged as an immensely desirable material across various applications beyond just telecommunication purposes, owing to its exceptional optical transmission properties, including remarkable tensile strength, dimensional stability, efficient thermal conductivity, and impressive heat resistance [2].

Other applications beyond telecommunication purposes include laser power delivery for medical, military, and manufacturing applications, sensing applications, lighting applications, and much more. At the forefront of these applications are silica-based fibers, primarily due to their ability to achieve the minimum optical loss of 0.15dB/km at visible wavelengths [3]. This critical characteristic has revolutionized data transmission efficiency and cemented silica fiber's position as the material of choice.

However, challenges emerge when operating within the IR light spectrum, where optical loss skyrockets to 800 dB/m, posing a significant obstacle for silica-based fibers where IR applications are needed [4]. There has been a growing trend toward exploring soft glasses such as heavy-metal oxide, fluoride, and chalcogenide to address this issue. These materials offer potential solutions to mitigate the elevated optical loss in the IR spectrum for silica-based fibers. While soft glasses have demonstrated promise in optical fiber applications in the IR spectrum, they still encounter signal loss compared to silica glass. Nonetheless, their unique characteristics, including a wide transmission range, high refractive index, and low phonon energy, set them apart and present intriguing opportunities for innovation and advancement within the industry [3-5].

The discovery in 1994 by Marshall et al. that microcrystals of ZBLAN can be suppressed in microgravity opened new avenues for achieving ZBLAN with minimal crystallization, leading to the realization of ultra-low attenuation rates i.e., 0.001dB/km [6].

Since then, extensive research has been conducted on ZBLAN, with a primary focus on microgravity conditions. Torres et al. [7] have determined that the inhibition of crystal growth in ZBLAN is predominantly influenced by heat transport mechanisms, specifically natural convection, and diffusion. These factors can be effectively suppressed within a temperature range of 360°C to 380°C. Additionally, Torres et al. [8] found that under normal gravity (1g), the degree of crystallinity increases with crystallization temperature between 360°C to 400°C. However, in a microgravity environment, the rate of crystallinity remains constant between 340°C and 400°C. This indicates the potential for suppressing ZBLAN crystallinity in microgravity conditions.

One of the common approaches to achieving microgravity is through the International Space Station (ISS). According to the data from 2021, the round-trip cost of transporting one kilogram of cargo to and from the ISS from Earth amounts to $60,000. Additionally, the expenditure associated with one hour of crew member time is $130,000 [9]. These substantial costs pose a significant barrier, potentially rendering ZBLAN fabrication prohibitively expensive in a microgravity environment. Moreover, as per the ISS, microgravity conditions typically exhibit an average vibration frequency ranging from 15 Hz to 50 Hz. This underscores the presence of the variable vibration frequency which plays a role in suppressing microcrystals within a microgravity environment [10].

To mitigate the high cost associated with the fabrication process, Michael et al. [11] proposed the implementation of a Low Earth Orbit (LEO) space station as a financially feasible alternative. While this approach may offer some reduction in expenses, it introduces a significant concern regarding carbon emissions. Commercial space manufacturing has the potential to substantially increase the carbon footprint of the space industry, thereby exacerbating climate change.

The influence of vibration frequency on ZBLAN crystallization has been largely overlooked in previous research. Shelby et al. [12] demonstrates that the rate of crystal growth in glass is influenced by mechanical vibration which affects atomic mobility within the material, impacting nucleation and crystal growth processes. Higher vibrational frequencies may accelerate nucleation and growth, contrary to the conditions observed in microgravity. Understanding this underlying parameter could significantly impact the fabrication process, potentially reducing costs and environmental impact by enabling ZBLAN production in 1g environments.

Therefore, this study aims to understand the influence of vibrations on the crystallization of ZBLAN under 1g condition. By elucidating the effects of vibrations on ZBLAN crystallization, future researchers and ZBLAN producers can gain a deeper understanding of how to regulate the fabrication process of ZBLAN. This understanding holds the promise of achieving the minimum attenuation rate of ZBLAN, thus facilitating its utilization in various applications. Similarly, the potential to reduce fabrication costs by a significant amount opens new avenues for the broader utilization of ZBLAN in the fiber optics industry. This not only addresses the current limitations associated with high fabrication costs but also stimulates further research and innovation in the field.

**Experimental Method**

To investigate the impact of vibration on the crystallization of ZBLAN a previously developed ZBLAN processing apparatus (developed by Torres et al. [13]) was retrofitted for this study and is shown in Figure 1.

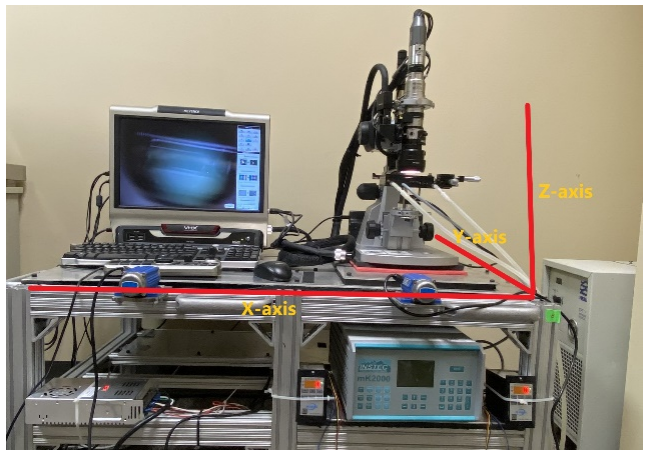

Figure 1: ZBLAN testing apparatus.

The modified ZBLAN testing system consisted of an 80/20 aluminum frame equipped with a Keyence VHX-2000 digital optical microscope capable of 100 to 1000X magnification using various microscopic imaging techniques. An Instec heated optical viewing stage was affixed to the microscope and provided precise temperature control from room temperature up to 500°C. The stage was integrated with an external temperature controller and a refrigerated water-circulation unit to maintain thermal stability and minimize radiative heat loss. The sample holder of the Instec stage was customized to securely accommodate the sealed silica ampoules containing ZBLAN specimens produced for this study as shown in Figure 2.

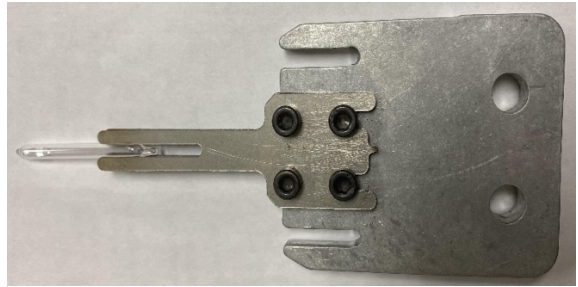

Figure 2: Ampoule holder along with ampoule.

To introduce vibration in addition to controlled heating, two miniature vibrating motors were mounted on opposite ends of the 80/20 frame. The motors operated independently at nominal speeds of 3000 RPM and 7000 RPM, corresponding to the low-speed and high-speed vibration modes, respectively. The low-speed motor was mounted directly beneath the microscope stage, while the high-speed motor was attached approximately 75 cm away from the stage to induce distinct vibration patterns. An accelerometer positioned directly below the heated stage (close enough for accurate sensing but far enough to avoid heat exposure) recorded the vibration frequencies along three orthogonal axes as shown in Figure 1.

Prior to the experiments, both motors were characterized using the accelerometer to establish vibration levels. Each control knob was divided into five discrete "levels" (H1-H5), representing incremental increases in vibration intensity. Each level was tested for 60 seconds and repeated three times per day over three different days to obtain reliable average vibration frequencies and amplitudes. These characterization results are presented in the results section.

The ZBLAN preform material used in this research was obtained from Fiber Labs and consisted of a core composition of 54%Zr - 27%Ba - 3%La - 3%Al -14%Na (ZBLAN) and a cladding composition of 50%Hf - 7%Ba - 5%La - 4%Al - 24%Na (HBLAN), consistent with that used in previous studies by Torres *et al.* [7, 8, 14]. The preforms had a cladding diameter of 1.0 ± 0.1 mm, a core diameter of 0.75 ± 0.1 mm, and were supplied in 15 cm lengths.

Sections approximately 3-5 mm in length were cut from the preform under a dry nitrogen atmosphere inside a glove box to minimize crystallization due to oxygen exposure. Each section was encapsulated within a sealed silica ampoule (3 mm diameter, 32 mm length) together with a small piece of silica glass acting as a separator to prevent ZBLAN to ZBLAN contact. The ampoules were evacuated to remove air and sealed on both ends to eliminate oxygen exposure. Each ampoule contained two identical ZBLAN samples to ensure repeatability. Approximately

200 ampoules were prepared by AG Scientific using consistent fabrication procedures. The final ampoule geometry is shown in Figure 3.

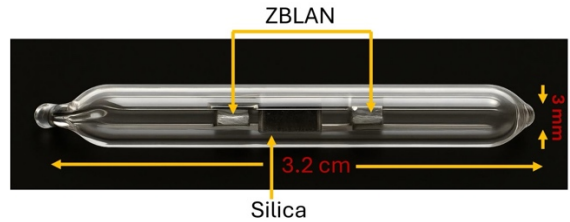

Figure 3: Ampoule containing two pieces of ZBLAN and a silica separator.

The possibility of intermittent thermal decoupling between the ZBLAN fiber and the inner wall of the silica ampoule during vibration is recognized as a potential source of temperature variation. Because the ampoules were evacuated, convective heat transfer was negligible, and any temporary loss of conductive contact could momentarily alter the local temperature experienced by the sample. Although the heating stage was well calibrated and maintained a stable external temperature, the exact temperature distribution within the ampoule during vibration could not be directly verified experimentally. A detailed thermal analysis using COMSOL Multiphysics was conducted to evaluate this effect, and the corresponding results, which quantify internal temperature gradients under vibrational conditions, will be presented in a forthcoming publication.

All ZBLAN samples used in both the thermal-only and vibration-assisted experiments were prepared from the same preform batch and underwent identical ampoule fabrication and sealing procedures. Each ampoule contained two samples of uniform composition and geometry to maintain consistency across all experimental conditions. Therefore, any variations observed between thermally treated and vibration-treated samples can be attributed solely to the applied experimental parameters rather than material inconsistencies.

The testing procedure followed a sequential and controlled protocol. All equipment including the microscope, heated stage controller, vibration motors, cooling unit, and data acquisition systems was powered on and allowed to stabilize. The cooling system water level was checked and adjusted as needed to ensure proper flow through the microscope assembly. The microscope was set to the desired magnification and focused for optimal imaging. The vibration frequency and heating stage temperature were then adjusted to the required test conditions and verified using the accelerometer and a thermocouple, respectively.

Once the stage temperature stabilized at the target value, the ZBLAN ampoule mounted in the holder (Figure 2) was placed onto the heated stage. Because direct temperature measurement of the ZBLAN sample was not feasible, the stage temperature was considered as the effective treatment temperature. Each thermal or vibration-assisted treatment was conducted for one minute. The treatment duration of one minute was selected to approximate the short microgravity exposure period (~20 s) typically experienced in parabolic flight experiments. This duration was sufficient to capture the onset of crystallization.

The time-temperature response of the system was previously characterized by Torres et al. [13] using an identical setup. Their measurements, obtained with a thermocouple inserted beside the ampoule, confirmed that the sample temperature approaches the set-point exponentially, reaching 400°C after approximately 3.5 minutes at a 400°C setting. This calibration validates the thermal response of the stage and supports the accuracy of the heating conditions employed in this study.

Following treatment, the ampoule holder was removed, and each ampoule was quenched to room temperature using a wet sponge. This quenching process was standardized across all samples to ensure uniform post-treatment conditions and to suppress crystallization during cooldown.

After quenching, the ampoules were carefully opened using a glass cutter, and the ZBLAN samples were extracted for microscopic and structural analysis. All samples were characterized using an Axia Chemi SEM located at the Shared Research Operations (SRO) Center at Texas State University. Backscattered electron imaging in low-vacuum mode was employed to obtain detailed morphological and microstructural information.

For cross-sectional imaging, ZBLAN pieces were mounted in epoxy resin prior to polishing. The epoxy mixture was prepared using Allied Epoxy Set Resin (145-20005) and Hardener (145-20010) in a 20:3 weight ratio. The mounted samples were cured for 72 hours to ensure complete hardening as shown in Figure 4.

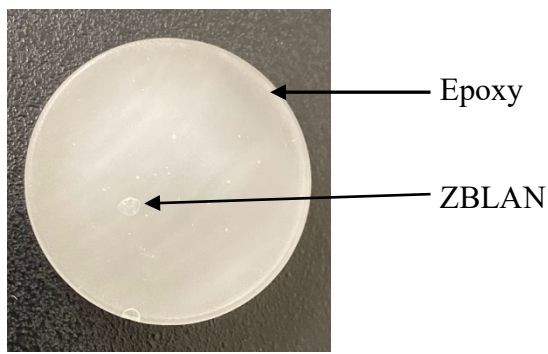

Figure 4: Epoxy mounted ZBLAN glass for SEM.

Mounted samples were polished using an Allied 230830 grinder-polisher. The grinding sequence began with 180-grit silicon carbide paper and progressed through 320, 600, and 1200 grit stages, each performed under water at 200 RPM for 2 minutes, with 90° rotation between grits to promote uniform material removal. Final polishing consisted of two stages: first with 6 μm polycrystalline diamond suspended in hexylene glycol on a polishing cloth at 200 RPM for 3 minutes, followed by a 1 μm diamond polish under identical conditions. All polishing steps were carried out under continuous water flow and low rotational speeds to minimize frictional heating and oxygen exposure. These precautions were taken to reduce the likelihood of surface-induced crystallization and to ensure that the microstructural features observed in SEM accurately represent the intrinsic structure of the samples.

**Results and Discussion**

Figure 5(a) represents the average recorded values for the high-speed vibrating motor. The vibration values for the five different levels increased along with an increase in the controller knob level, which is as expected. It is also observed, that at level 1, all three axes' values provide the same vibration frequency, at approximately 30 Hz. Increasing the knob to level 2 results in an average frequency of approximately 50 Hz, with some disparity between the three axes. Similarly, level 3 had an increase in vibration frequency with an average value of approximately 80 Hz, with some disparity coming from the x-axis. Levels 4 and 5 had the most disparity between the three axes, but it is observed that there is a marginal average difference between levels 3, 4, and 5. All three levels (3, 4, and 5) have an approximately average of 80 Hz. The disparity between the three axes at this high frequency is likely due to how the testing apparatus is assembled along with how the individual components are affixed to the testing apparatus, especially in the z-axis. In Figure 1, the z-axis is the vertical axis, whereas the x and y axes are horizontal and orthogonal to each other. Also, there exists a red and black colored piece of foam that is situated between the microscope and the testing apparatus frame. This was initially incorporated into the design from previous experimentation to dampen the vibrations from the parabolic aircraft. Therefore, as the motors are turned on, there is some dampening occurring due to the foam, which is likely causing the variability in the z-axes.

Similarly, Figure 5(b) shows the average reported values of the low-frequency vibrating motor, which is affixed to the testing apparatus in closer proximity to the heated stage. The mounting points for the two motors were selected based on the anticipated vibration frequency output, such that the higher frequency motor would induce too much vibration if affixed close to the heated stage, therefore, it was mounted further away from the heated stage and the lower frequency motor was mounted closer.

As observed in Figure 5(b), the low-frequency vibrating motor produces less than 5 Hz of vibration, on average, at the level 1 position of the controller knob. However, turning the knob to level 2 drastically increases the vibration to approximately 35 Hz, with no variability in the three axes. Turning the controller knob to level three increases the vibration to approximately 45 Hz, on average, with minor variability in the three axes. Lastly, increasing the controller knob on the low-frequency vibration motor results in 50 Hz for both levels 5 and 6, with virtually no variability in the three axes. This result makes sense as the power output of the low-frequency motor at 3000 RPM should output approximately 50 Hz of vibration.

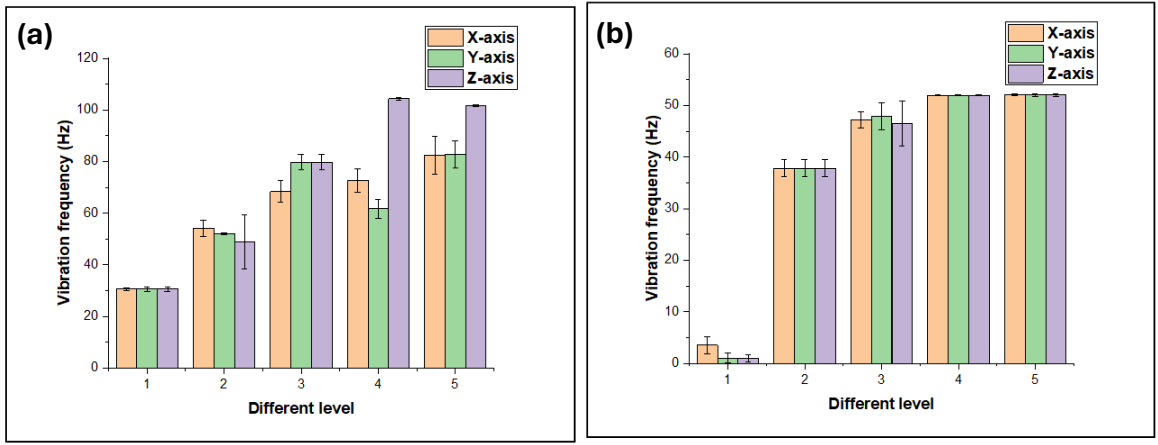

Figure 5: Vibration frequency value in five different levels of (a) high-speed vibrating motor, and (b) low-speed vibrating motor.

Similarly, the high-frequency motor produces 7000 RPM, which should output approximately 116 Hz of vibration. As seen in Figure 5(a) the high-frequency motor at levels 4 and 5 was able to produce nearly that value, with the z-axis producing values closer to 116 Hz.

Overall, this characterization was able to demonstrate the vibration frequency that is producible with the two vibrating motors affixed to the ZBLAN vibration testing apparatus in their current configuration. The results show that the low-frequency vibrating motor can produce a vibration frequency ranging from approximately 5-50 Hz and the high-frequency vibrating motor can produce a vibration frequency of 30-80 Hz.

To verify that the two vibrating motors operated within their designed electrical capacities, a power consumption analysis was performed using a PN2500 Electricity Usage Monitor. The device was connected in series between the power source and each motor, allowing real-time measurement of input power. Each motor was tested independently, and power readings were recorded across all vibration levels. The baseline power draw of the system, measured prior to motor activation, was 4.68 W.

Figure 6 depicts the power consumption of a low-speed and high-speed vibrating motor. For the low-speed vibrating motor with the increase in level from one to three, the rate of power consumption also increases from 9 watts (W) to 11.85W. However, beyond level three, power consumption does not increase as before. This is because the power output of the low-speed motor is limited to 3000 RPM, and by reaching level three, it has already attained its maximum capacity.

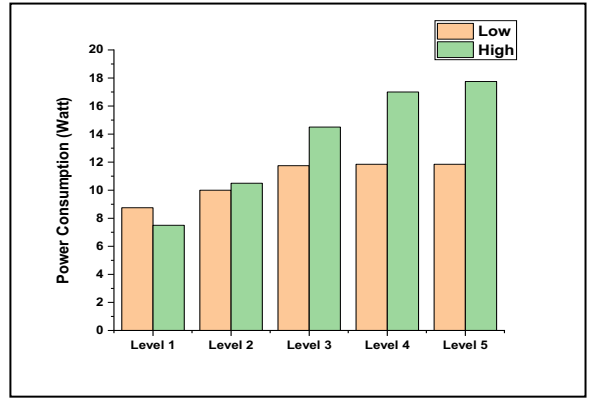

Figure 6: Power Consumption of low-speed and high-speed vibrating motors.

Similarly, for the high-speed vibrating motor, at level one the power consumption is approximately 7.5W, gradually increasing to 17W by level four. However, at level five, power consumption does not exhibit a comparable rate of increase to that observed in the preceding levels. This deviation is attributed to the high-speed motor's capacity of 7000 RPM; by reaching level five it has already reached its capacity.

Following the characterization of the vibrating motors, ZBLAN was tested at each level of vibration frequency provided by high vibrating motor and level 2, 3, and 4 provided by low vibrating motor. Level 1 and 5 of low-speed vibrating motor were not included as level 1 generated the negligible amount of vibration frequency and level 5 generated the same as level 4. The temperature range was chosen from 250ºC to 400ºC considering ZBLAN's glass transition temperature (Tg) and glass crystallization temperature (Tx) and known crystallization behaviors as reported by Torres et al. [7,8,13,14].

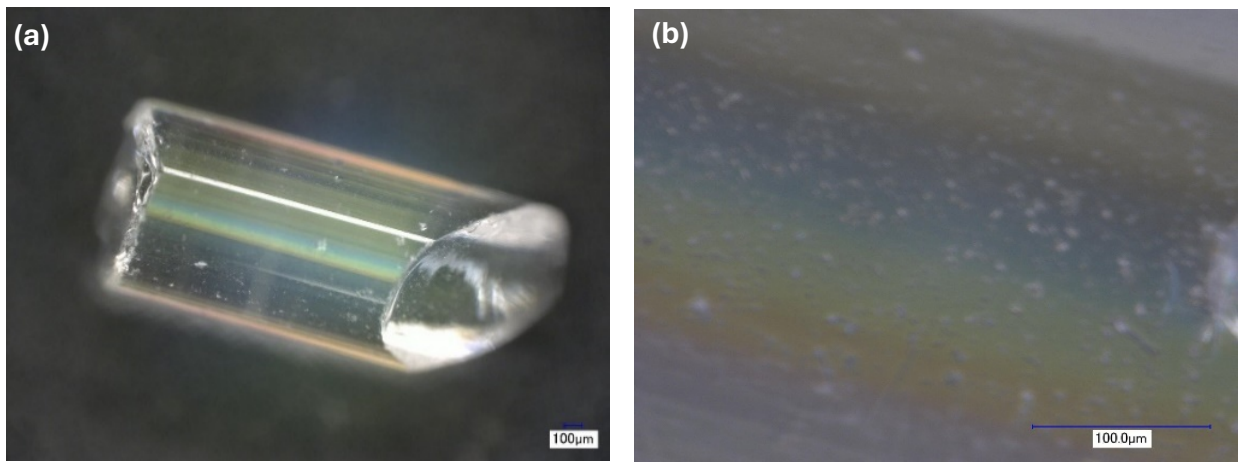

Figure 7: Microscopic images of ZBLAN control samples with a magnification of (a) 100, (b) 1000.

Figure 7 shows the microscopic images at varying magnification of as-received ZBLAN samples that have not been subjected to any heat treatment or vibration. These images were taken to provide baseline images for comparison purposes with the treated ZBLAN samples.

When observed in the control samples, the material exhibits relative clarity, with very small inclusions and/or surface features on the order of μm. It is difficult to ascertain whether these inclusions are on the surface of the sample or within the sample due to the nature of the microscope.

The minor inclusions could also be dust particles and the large inclusion in

Figure 7(b) could be a defect from production or handling. Not being able to determine the exact nature of the visible features is one of the major limitations of optical microscopy, which will be addressed by further characterization techniques such as SEM, AFM, etc.

Similarly, Figure 8 shows the microscopic images of ZBLAN treating them only with varied temperatures. Each ampoule was treated at varying temperature for 1 minute of duration and was immediately quenched using wet sponge as explained above. This treatment was also done to provide a baseline for comparison purposes of ZBLAN treated with both vibration frequencies and temperature. Initial observation shows that when the temperature is initially raised from 250°C to 330°C, the ZBLAN retains its original structure and remains transparent.

However, at 340°C, incipient crystallization began within the ZBLAN. This continues until it reaches 350°C. At 360°C and 370°C the ZBLAN has partially crystallized, where transition from incipient to partial crystallization occurs. As the temperature is further increased up to 400°C ZBLAN remains in a predominantly crystalline and opaque state.

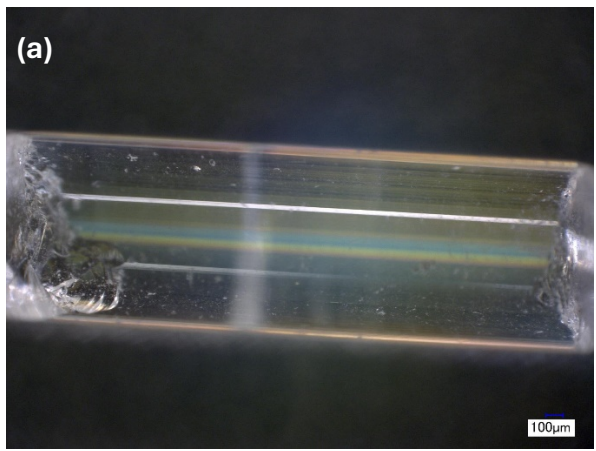
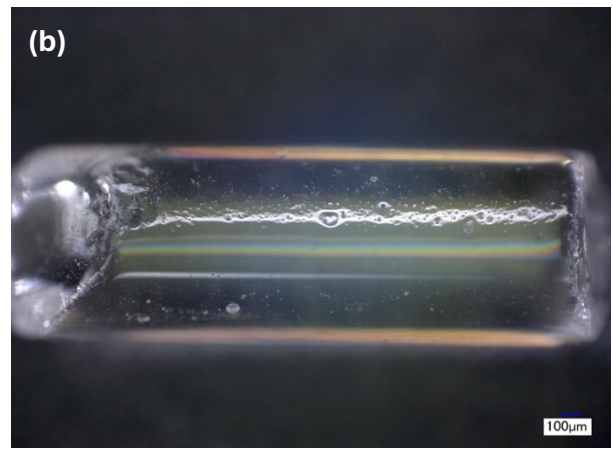

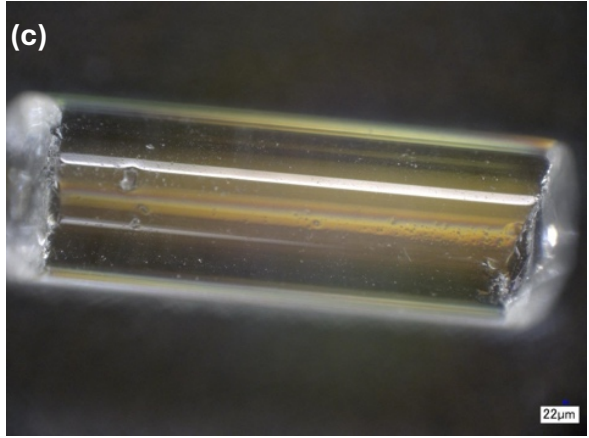

(c)

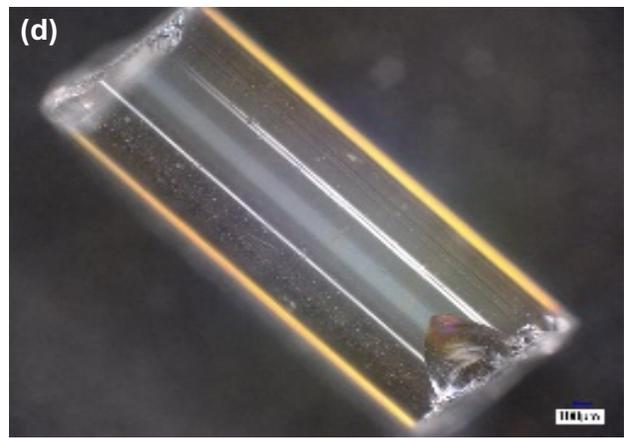

(d)

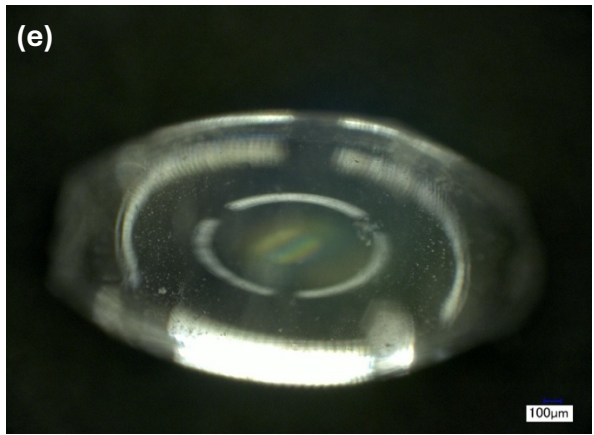

(e)

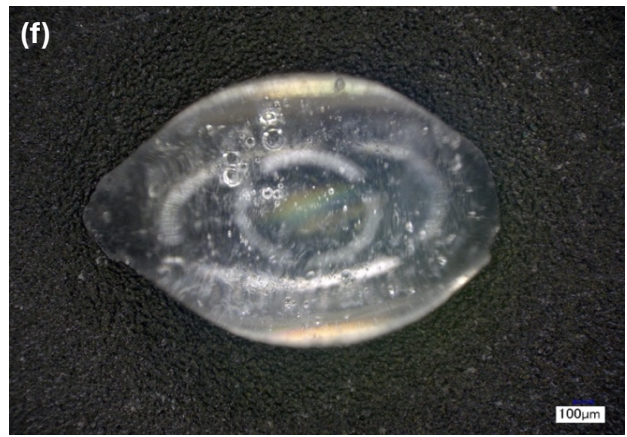

(f)

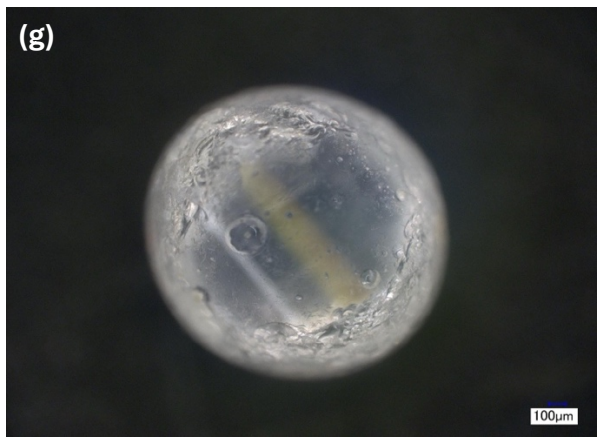

(g)

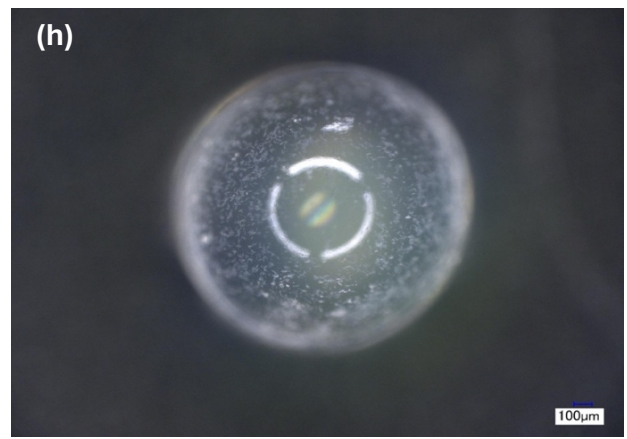

(h)

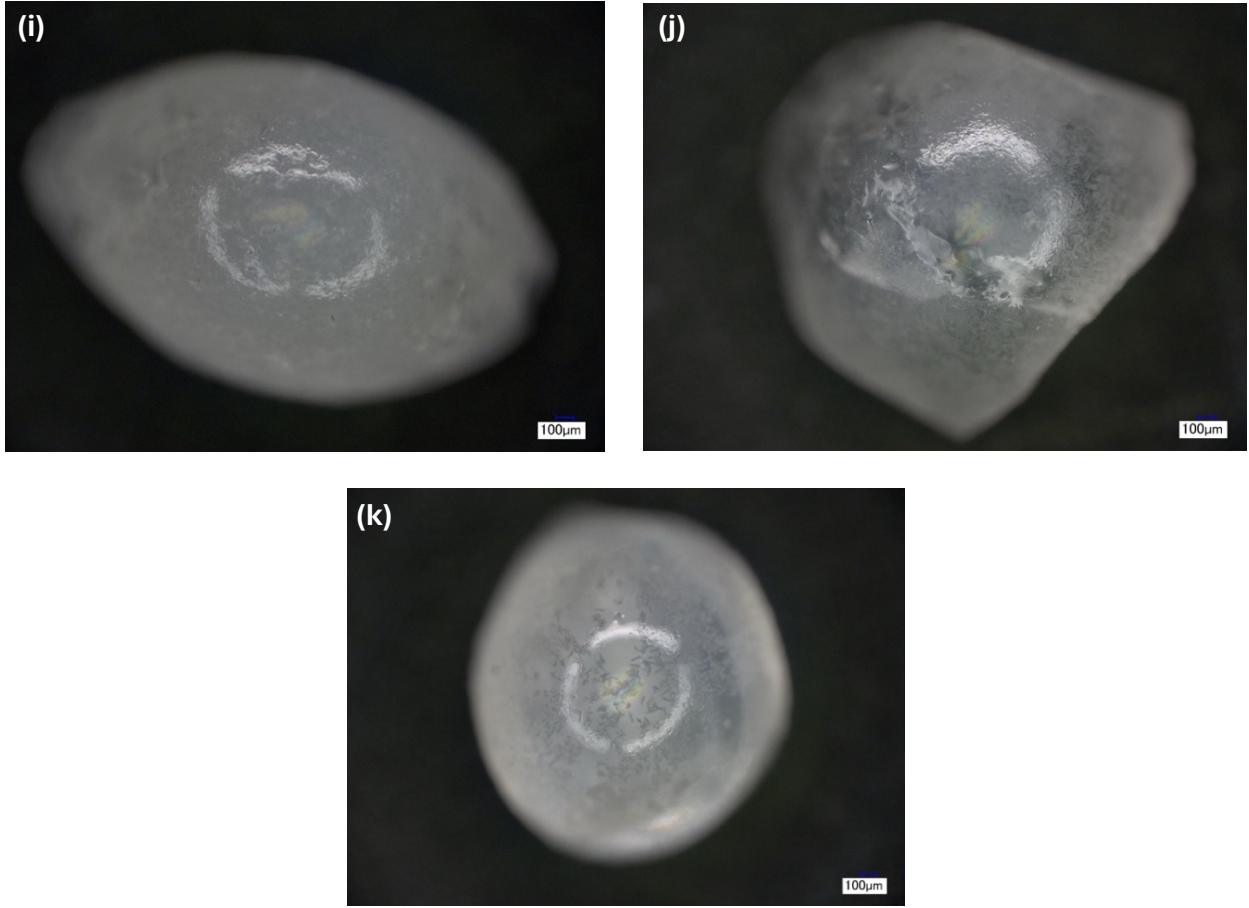

Figure 8: Microscopic images of ZBLAN after testing with varied temperature (a) 250°C, (b)300°C, (c) 320°C, (d) 330°C, (e) 340°C, (f) 350°C, (g) 360°C, (h) 370°C, (i) 380°C, (j) 390°C, (k) 400°C

Figure 9 illustrates the microscopic images of ZBLAN following one minute at temperatures ranging from 250°C to 400°C. These conditions were combined with three different low-vibration levels, as discussed in Figure 5(b). In level two, with a vibration of 37.86 Hz in all three axes, the ZBLAN remains uncrystallized until it reaches 320°C. Crystallization begins at 330°C onwards, continuing up to 400°C.

Similarly, in level three, with vibration of 47.15 Hz, 47.90 Hz, and 46.52 Hz in the X, Y, and Z axis respectively, incipient crystallization starts occurring at 330°C. However, extensive crystallization occurs after reaching 360°C, continuing up to 400°C. Likewise, in level four, with a vibration of 51.98 Hz in all three axes, crystallization begins at 350°C. Crystallization intensifies beyond 360°C, continuing up to 400°C.

In all three different vibration levels, incipient crystallization begins at 340°C and 350°C and becomes more pronounced near 360°C. However, the low-speed motor does not yield varying results compared to the high-speed motor. The current results lead to the conclusion that the vibration of the low-speed motor is insufficient to alter the crystallization structure of ZBLAN at different temperatures.

| Temp./Freq | Low Level 2<br>X-axis:37.86 Hz<br>Y-axis:37.86 Hz<br>Z-axis:37.86 Hz | Low Level 3<br>X-axis:47.15 Hz<br>Y-axis:47.90 Hz<br>Z-axis:46.52 Hz | Low Level 4<br>X-axis:51.98 Hz<br>Y-axis:51.98 Hz<br>Z-axis:51.98 Hz |
|---|---|---|---|
| **250°C** | 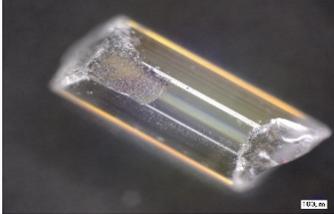 | 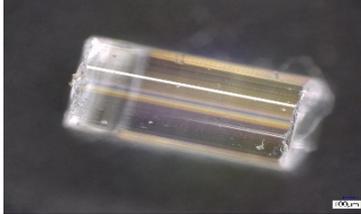 | 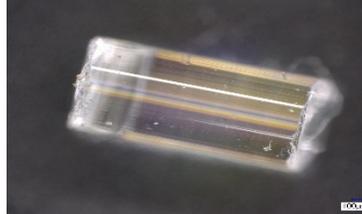 |
| **300°C** | 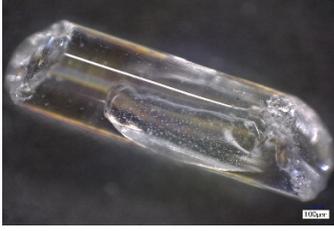 | 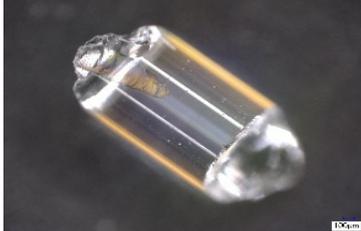 | 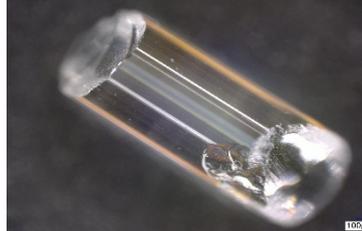 |
| **320°C** | 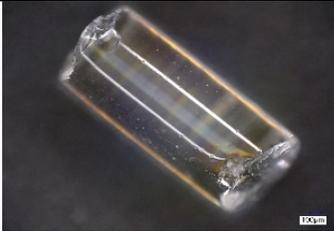 | 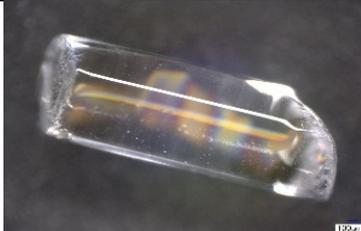 | 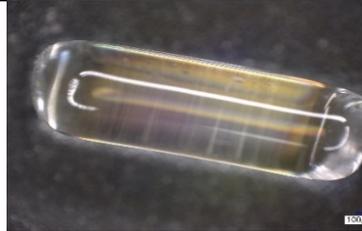 |
| **330°C** | 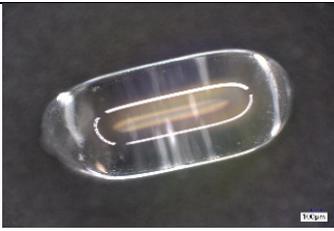 | 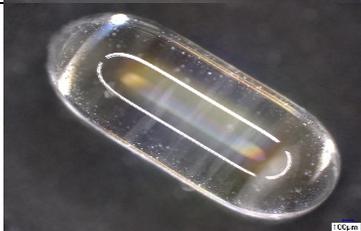 | 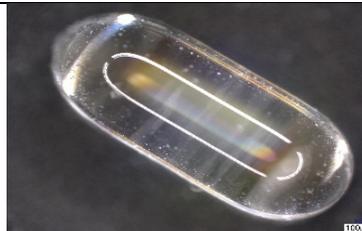 |
| **340°C** | 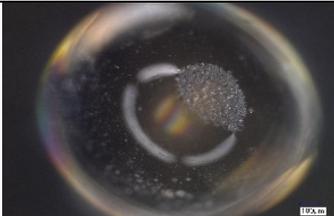 | 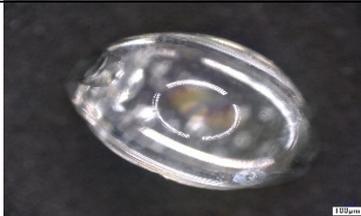 | 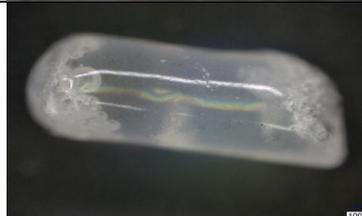 |
| **350°C** | 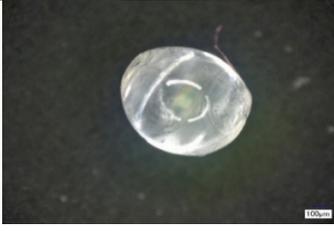 | 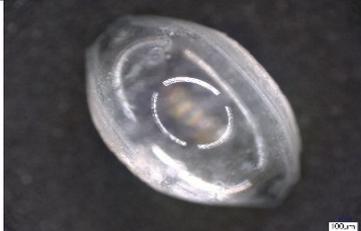 | 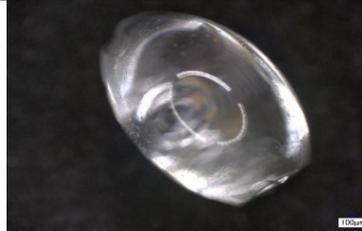 |

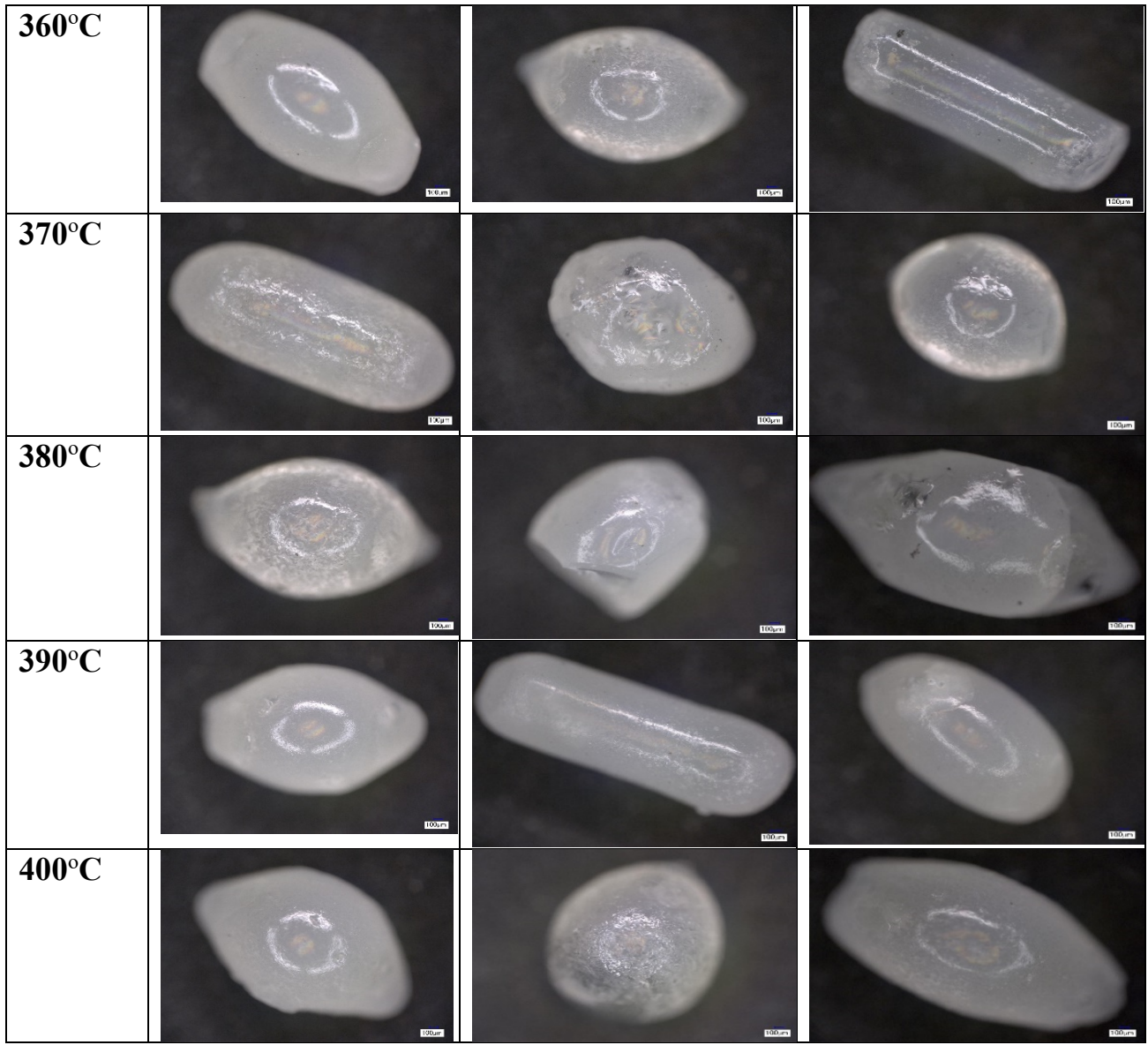

Figure 9: Microscopic images of ZBLAN after testing after varied temperatures and vibration frequencies with a low-speed vibrating motor.

Similarly, Figure 10 depicts the microscopic images of ZBLAN following for one minute under various temperatures ranging from 250°C to 400°C These conditions were combined with five different high-vibration levels, which as discussed in Figure 5(a). In the first level, characterized by a vibration of approximately 30Hz in all three axes, the ZBLAN remains unchanged in an amorphous state until it reaches 330°C. At 340°C, the incipiently crystallization begins, while extensive crystallization is observed only from 360°C onwards, continuing up to 400°C.

| Temp/Freq. | High Level-1<br>X-axis:30.58 Hz<br>Y-axis:30.57 Hz<br>Z-axis:30.57 Hz | High Level-2<br>X-axis:54.08 Hz<br>Y-axis:52.07 Hz<br>Z-axis:48.48 Hz | High Level-3<br>X-axis:69.39 Hz<br>Y-axis:79.72 Hz<br>Z-axis:79.72 Hz | High Level-4<br>X-axis:72.61 Hz<br>Y-axis:61.66 Hz<br>Z-axis:104.29 Hz | High Level-5<br>X-axis:82.40 Hz<br>Y-axis:82.72 Hz<br>Z-axis:101.62 Hz |
|---|---|---|---|---|---|
| 250°C | 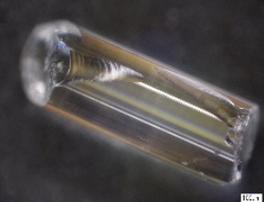 | 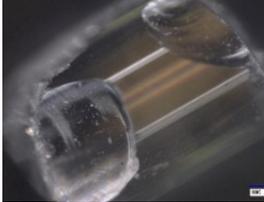 | 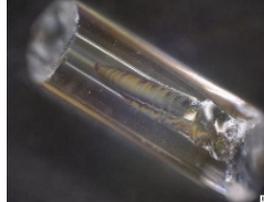 | 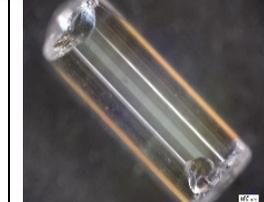 | 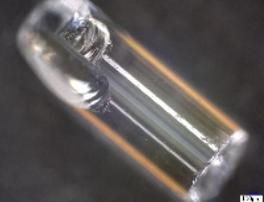 |
| 300°C | 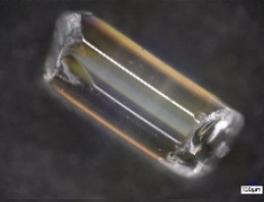 | 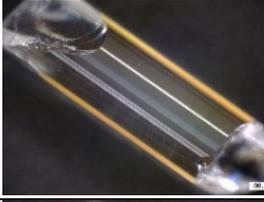 | 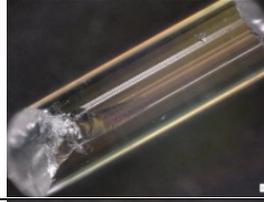 | 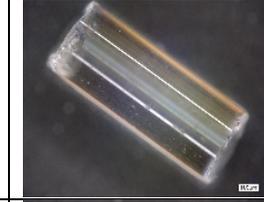 | 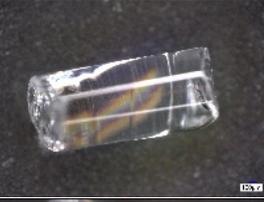 |
| 320°C | 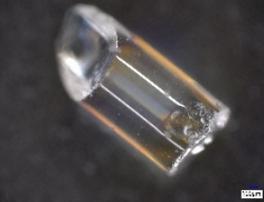 | 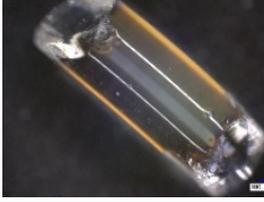 | 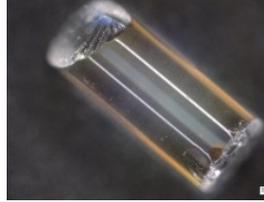 | 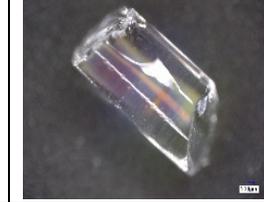 | 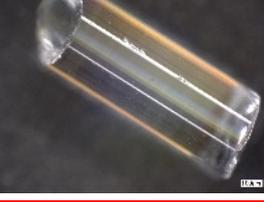 |
| 330°C | 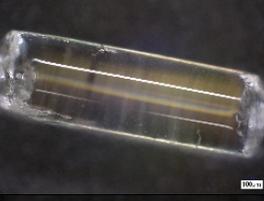 | 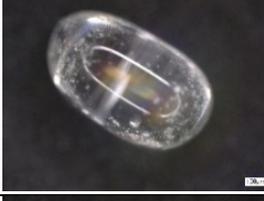 | 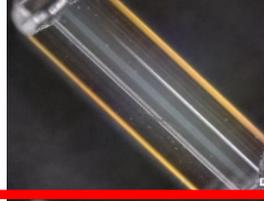 | 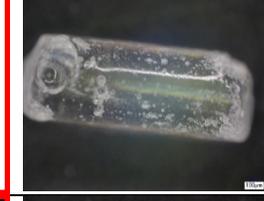 | 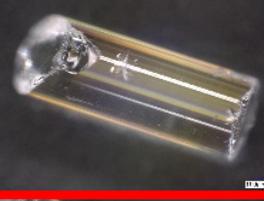 |
| 340°C | 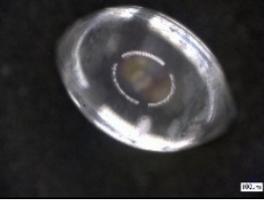 | 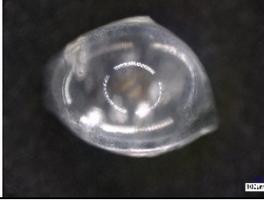 | 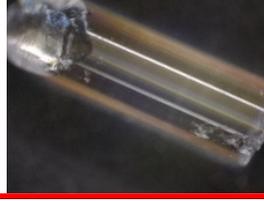 | 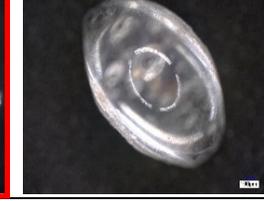 | 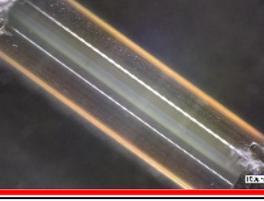 |

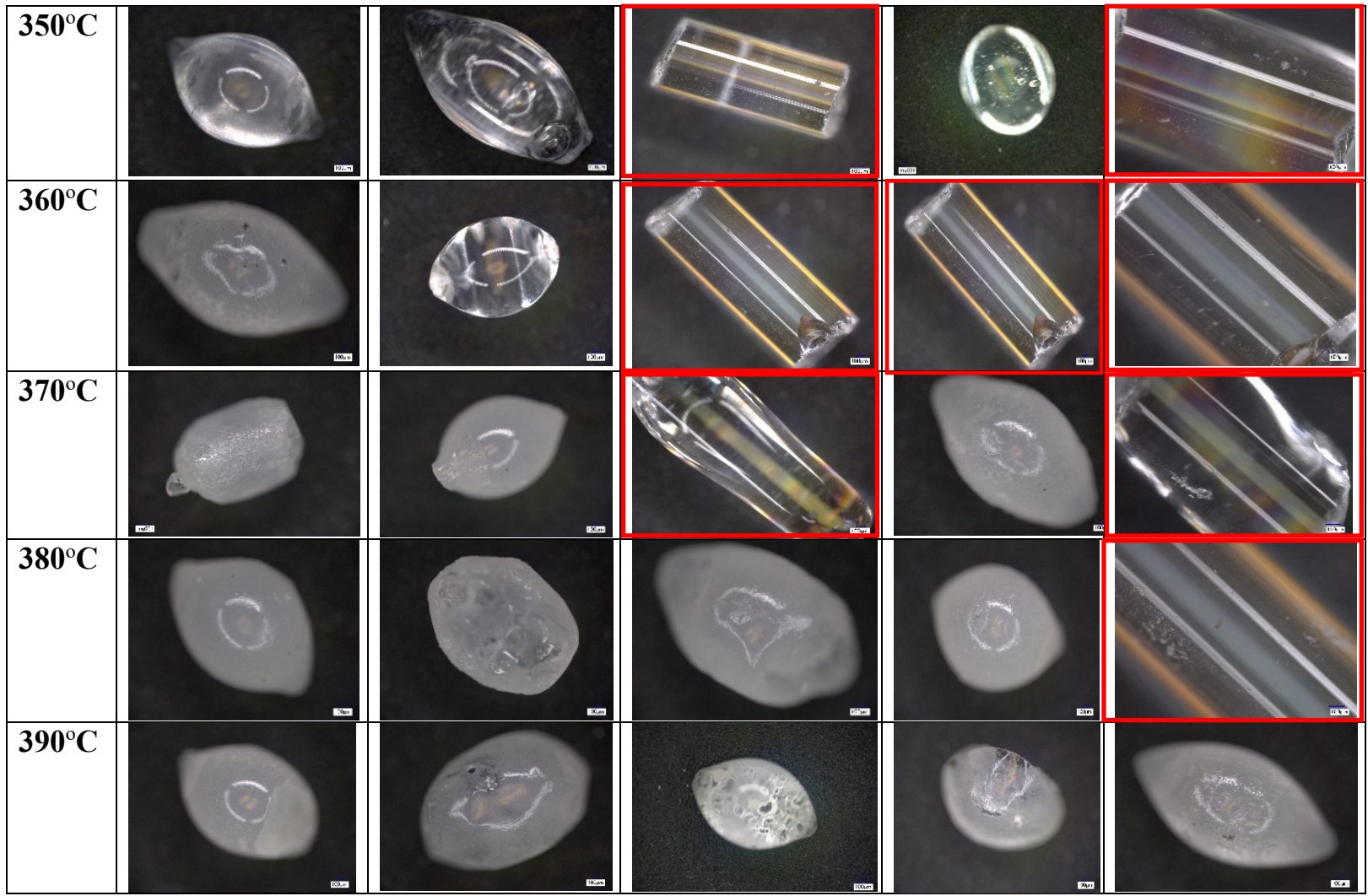

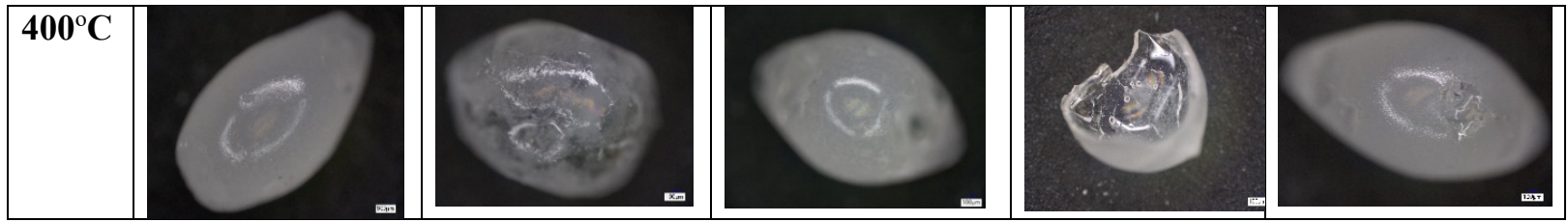

Figure 10: Microscopic images of ZBLAN after testing with varied temperatures and vibration frequencies with a high-speed vibrating motor.

Likewise, at level two with vibration of 54.08 Hz, 52.07 Hz, and 48.48 Hz in the X, Y, and Z axes respectively, incipient crystallization starts occurring at 330°C. However, crystallization becomes well developed after 370°C, continuing up to 400°C. Similarly, at level three with vibration of 69.39 Hz, 79.72 Hz, and 79.72 Hz in the X, Y, and Z axes respectively, crystallization solely at 380°C.

At level four, which has vibration of 72.61 Hz, 61.66 Hz, and 104.29 Hz in the X, Y, and Z axes respectively, incipient crystallization occurred at 330°C and 340°C. However, at 350°C and 360°C, the ZBLAN remains unchanged. Pronounced crystallization becomes evident from 370°C onwards, persists up to 400°C. Lastly, at level five, with vibration of 82.40 Hz, 82.72 Hz, and 101.62 Hz in the X, Y, and Z axes respectively, crystallization commences only at 390°C and 400°C.

Although the overall trend indicates that vibration enhances crystallization, several inconsistencies were observed at higher vibration levels. Specifically, as highlighted by the red boxes in Figure 10, the samples treated at vibration levels H3, H4, and H5 did not exhibit consistent morphological or optical changes at certain temperatures where crystallization was otherwise expected. This irregular behavior is likely attributed to insufficient thermal exposure caused by increased jostling of the samples within the evacuated ampoule at elevated vibration intensities, which temporarily reduced thermal contact with the heated surface. Video analysis supports this interpretation, revealing that these samples experienced more pronounced movement and occasionally shifted out of the microscope's field of view. Based on these observations, a hypothesis was formulated that the jostling effect impedes effective heat transfer to the ZBLAN sample during vibration. To evaluate this hypothesis, a detailed thermal analysis was conducted using COMSOL Multiphysics, and the corresponding results, along with the redesign of the testing apparatus and subsequent retesting of the inconsistent samples, will be presented in a forthcoming publication.

Despite these observed inconsistencies, the collective results from the temperature only, low-vibration, and high-vibration experiments demonstrate that both temperature and vibration significantly influence the crystallization behavior of ZBLAN. Crystallization initiates between 330°C and 350°C and becomes progressively more pronounced with increasing temperature. When comparing samples subjected to identical thermal conditions, those exposed to higher vibration levels exhibit earlier onset and greater crystallization extent than those treated under low-vibration conditions. This indicates that mechanical vibration enhances atomic mobility and promotes nucleation, thereby accelerating the transition from the amorphous to the crystalline state.

To further investigate the microstructural effects of vibration on ZBLAN crystallization, scanning electron microscopy (SEM) was employed to characterize the morphology and distribution of crystalline phases formed under different thermal and vibrational conditions. SEM provides high-resolution visualization of surface features, enabling correlation between the optical observations of cloudiness and the underlying crystalline development. This analysis was performed on multiple samples to ensure reproducibility and consistency across experimental conditions. Representative micrographs are presented below to illustrate the characteristic structural changes induced by vibration and to support the interpretation of nucleation and growth behavior discussed earlier.

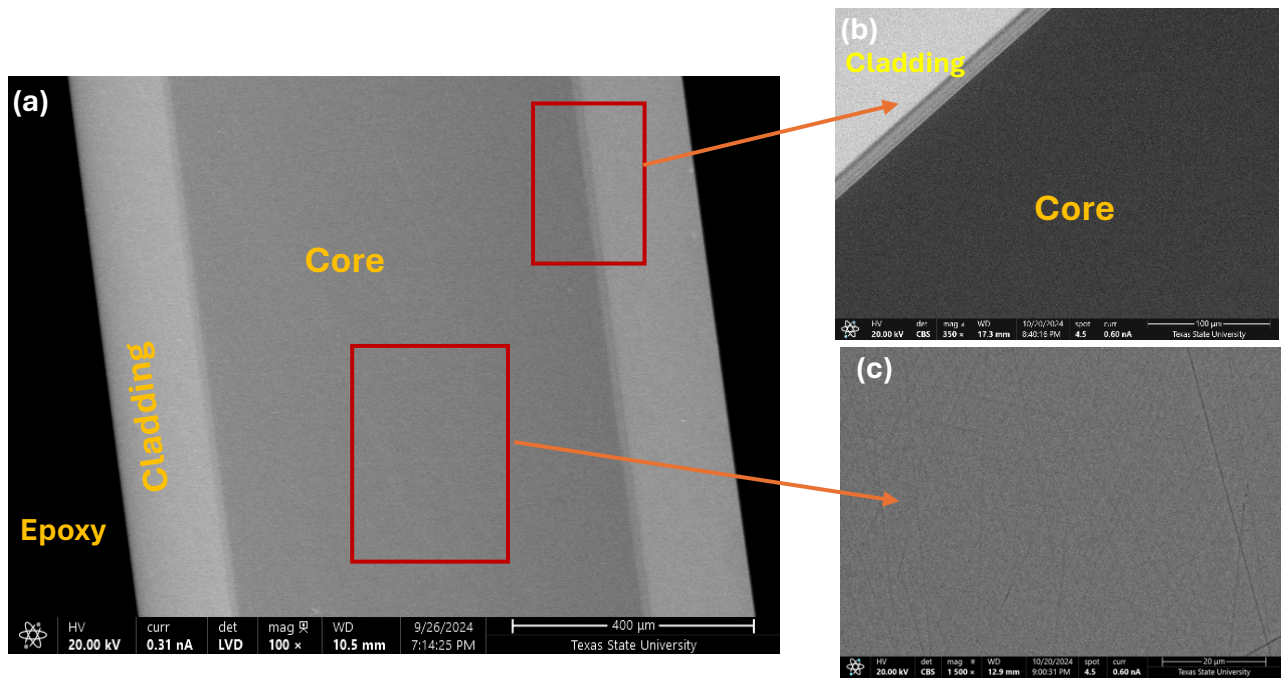

Figure 11: SEM images of ZBLAN showing core and cladding layer.

The backscattered SEM image of the ZBLAN preform, as provided by Fiber Labs, is shown in Figure 11. This image reveals two distinct layers: the cladding and the core. The outer cladding layer has an average width of 174.5 μm, while the core has an average width of 653.74 μm. Figure 11(b) presents a magnified view encompassing both the core and cladding, while Figure 11(c) provides a closer look at the core region of ZBLAN, as shown in Figure 11(a).

The contrast in color between the core and cladding layers is attributed to their compositional differences. Specifically, the cladding layer contains hafnium (Hf) in place of zirconium (Zr), which is found in the core. The brighter appearance of the cladding layer in the SEM image is due to the difference in atomic number between hafnium and zirconium. Hafnium, with an atomic number of 72, backscatters more electrons compared to zirconium, which has an atomic number of 40. This higher atomic number results in a more intense backscattered electron signal, making the cladding appear brighter in SEM imaging [15,16]. Additionally, hafnium's higher density compared to zirconium further contributes to this brightness difference, as denser materials tend to scatter more electrons, enhancing the contrast observed in the SEM images [16].

A thin intermediate layer is also observed between the core and cladding regions, forming a narrow transition zone. This feature likely represents a compositional interdiffusion layer created during the preform fabrication process, where limited fluoride ion diffusion occurs between the ZBLAN and HBLAN glasses at elevated co-drawing temperatures. The presence of this layer indicates effective bonding and compositional continuity between the two regions, confirming the structural integrity and high manufacturing quality of the preform prior to any heat or vibration treatment.

The SEM images of ZBLAN after treating with different set of temperature are depicted in the Figure 12. At 320ºC and 340ºC, no surface crystallization were observed, where the two distinct layers, core and cladding remained clearly visible as shown in Figure 12(a) and Figure 12(b) respectively.

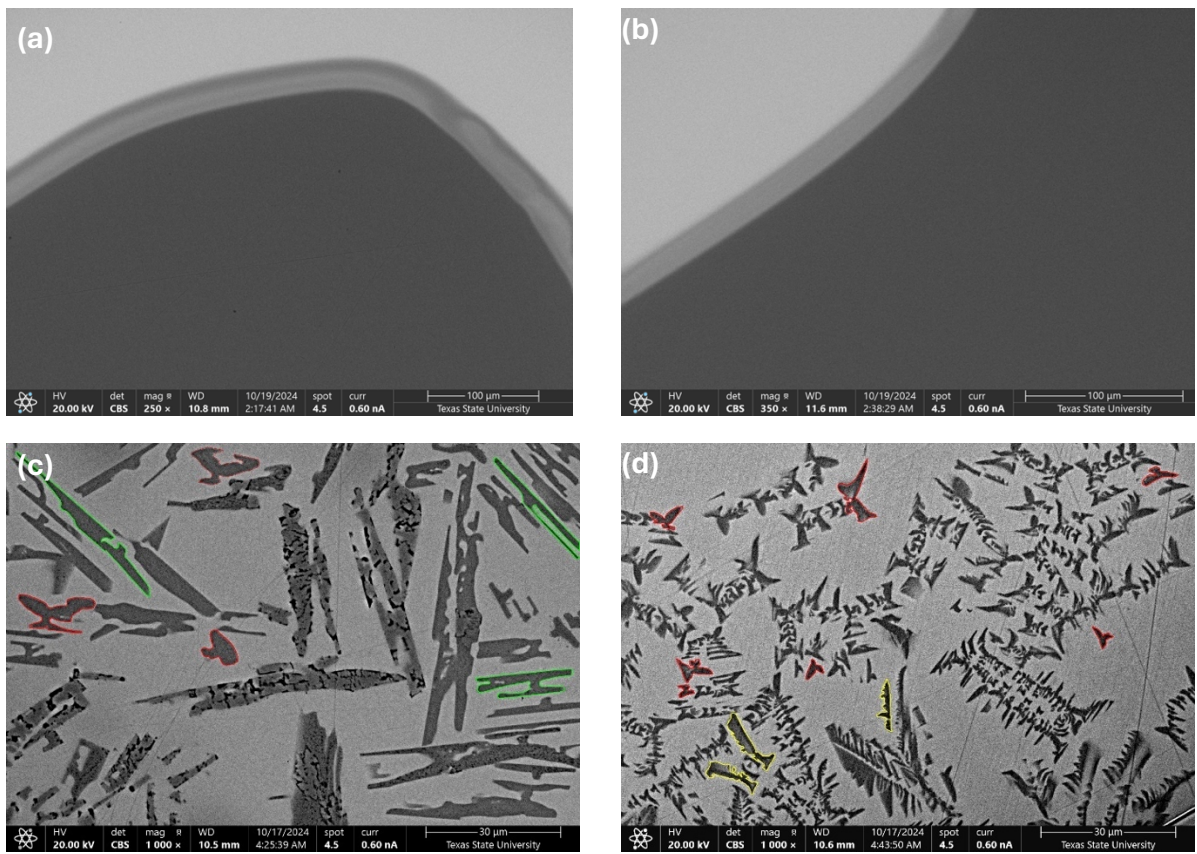

Figure 12: SEM images of ZBLAN while treating with different temperatures, (a) 320°C, (b) 340°C, (c) 390°C, and (d) 400°C.

At a temperature of 390°C, multiple crystals appeared on the surface of ZBLAN, as depicted in Figure 12(c). The surface exhibited two distinct types of crystals: dark needle-shaped crystals (highlighted in green) and bow-tie-shaped crystals (highlighted in red). These crystals collectively occupied 33.92% of the SEM image area, which measured 3641.192 µm², with the largest crystal measuring 566.934 µm².

At a temperature of 400°C, distinctive crystals of varying shapes and sizes formed on the surface of ZBLAN, as illustrated in Figure 12(d). Two prominent morphologies were observed: bow-tie-shaped crystals (highlighted in red) and feather-like crystals (highlighted in yellow). These crystals collectively covered 22.04% of the SEM image area, which measured 2370.257 µm². Among the observed crystals, the largest had an area of 113.031 µm².

The sample treated at 380ºC is discussed separately, as it exhibited distinct and heterogeneous crystallization behavior compared to other temperature conditions. As shown in Figure 13(a) clusters of crystals forming star-like aggregates were observed, with needle-shaped crystals forming along the outer edges. Within these clusters in Figure 13(b), the average crystal size was 1.973 µm², covering a total area of 98.673 µm². On the opposite surface, Figure 13(c), needle-shaped crystals (highlighted in green) dominated, averaging 7.027 µm² and occupying 456.762 µm² (26.53 % of the imaged area). Additionally, bow-tie- and feather-like crystals in Figure 13(d) were observed, covering 2858.877 µm² (26.53 % of the area).

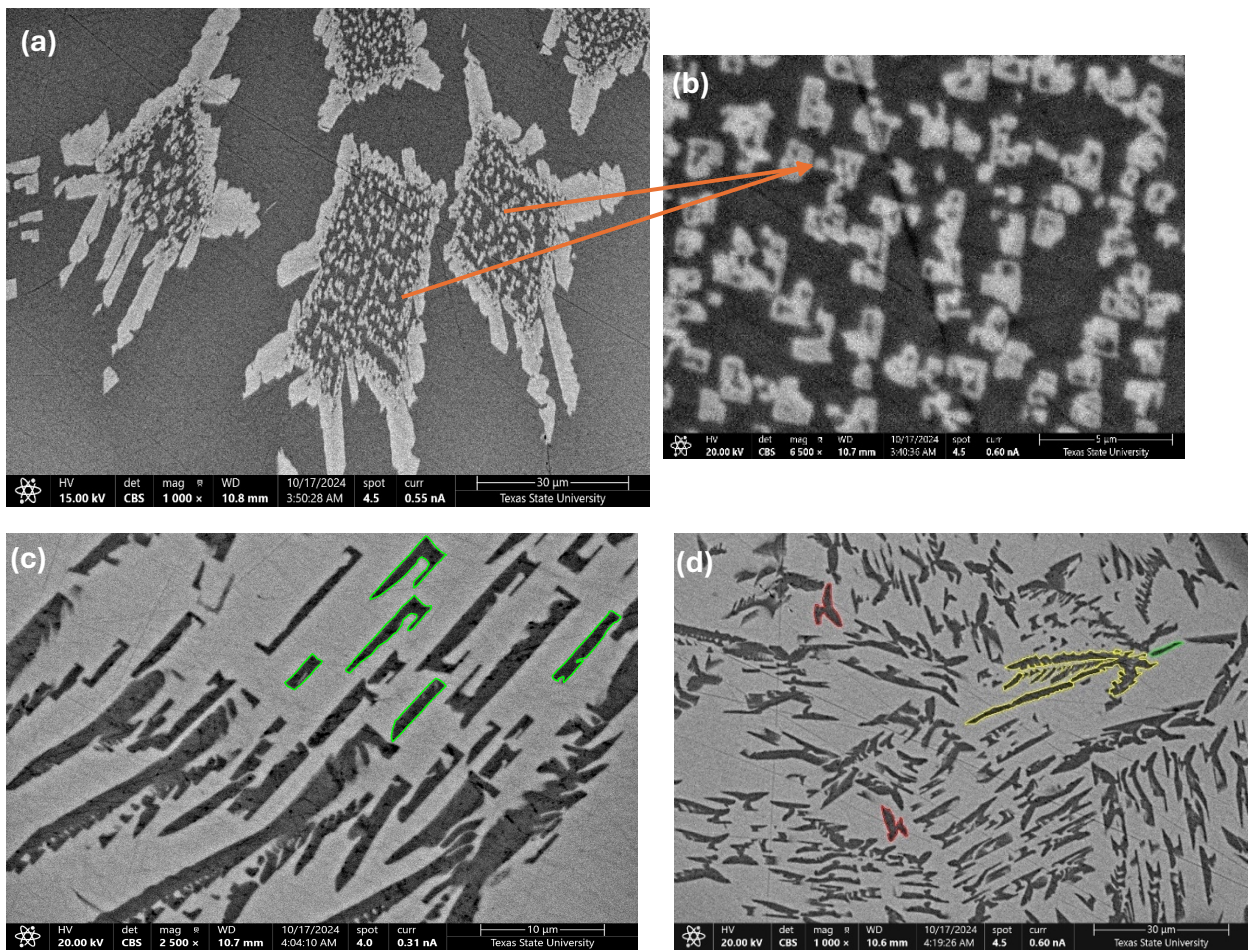

Figure 13: SEM images of ZBLAN treated at 380°C.

The diversity of morphologies and surface coverage at 380°C suggests that this temperature represents a transitional regime in the crystallization of ZBLAN, where multiple nucleation mechanisms operate simultaneously. At this stage, atomic diffusion within the glass matrix becomes sufficient to initiate crystallization, yet localized temperature gradients and compositional variations lead to the coexistence of needle-shaped, bow-tie, and feather-like crystals. This combination of features indicates the competition between homogeneous and heterogeneous nucleation processes, marking 380°C as the critical threshold for the onset of extensive crystallization in ZBLAN.

Building on the thermal only observations, SEM analyses were also performed on ZBLAN samples subjected to combined thermal and vibrational treatments to clarify the influence of mechanical excitation on crystallization. The objective was to examine how vibration intensity, applied through both low and high-speed motors, affects crystal nucleation, growth, and morphology. Representative results for selected temperature and vibration conditions are discussed below.

Figure 14 depicts the SEM images of ZBLAN treated at a temperature of 390°C and a vibration of L2. The crystals were observed forming along the edges in a circular arrangement, as highlighted in Figure 14(a). The contrast between white and black in the image arises from the difference in elemental composition, as previously discussed.

Figure 14(b) provides a magnified view of the darker upper region of Figure 14(a), showcasing crystal clusters that merge to form a star-shaped crystal with needle-like projections pointing outward. Figure 14(c) focuses on a small section of Figure 14(a), where bow-tie-shaped crystals and feather-like crystal morphologies dominate. These structures account for 43.85% of the total SEM image area, covering an impressive 3270.8 μm². This highlights the extensive crystallization and unique morphological diversity induced by the specific thermal and vibrational treatment.

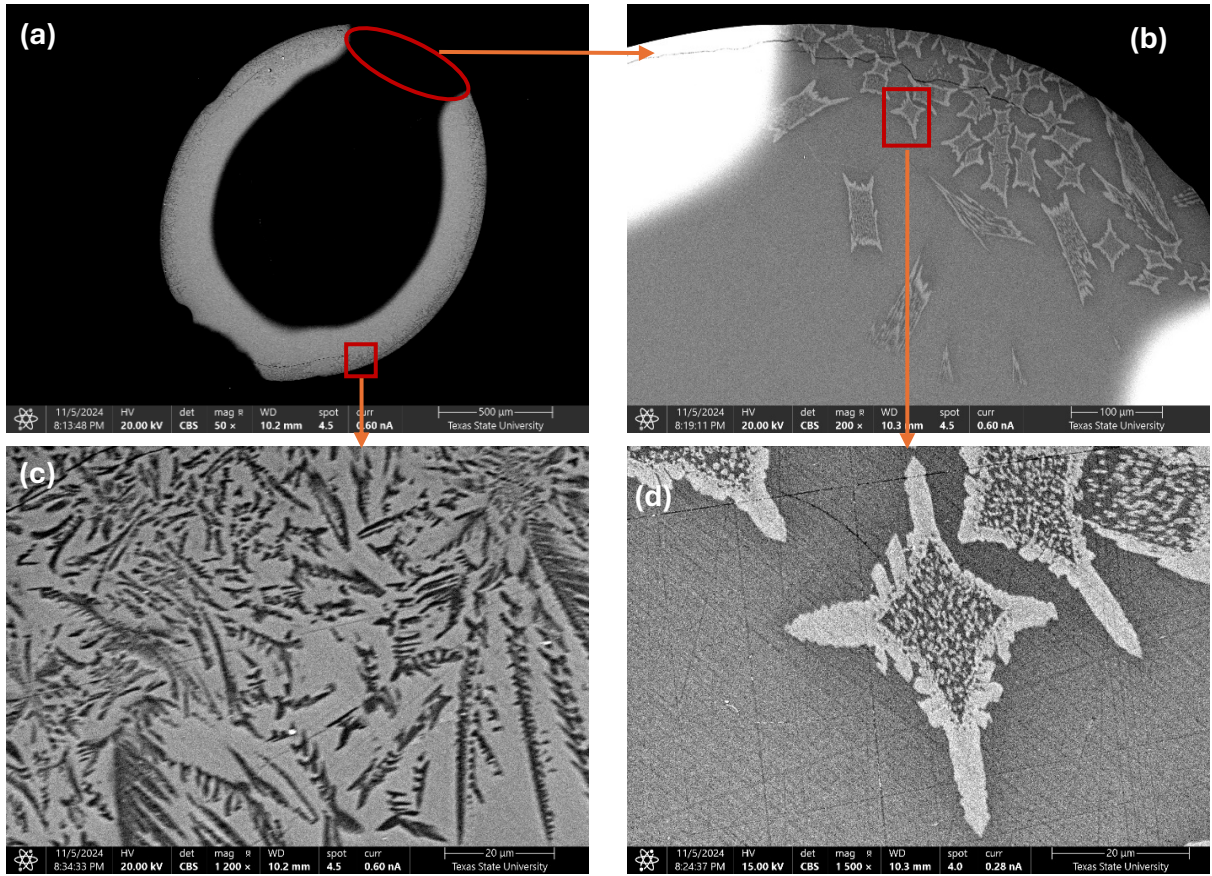

Figure 14: SEM images of ZBLAN treated at vibration frequency of level L2 and temperature of 390°C.

At a vibration of level H1 and a temperature of 360°C, multiple crystals were observed, as shown in Figure 15. In Figure 15(a), crystals are visible across the entire surface of the ZBLAN, including both edges and the corners. A magnified portion of this surface is shown in Figure 15(b), with a further magnification provided in Figure 15(c). In Figure 15(b), the crystals cover 22.74% of the SEM image area, corresponding to a total crystal area of 5816.96 μm², with the largest individual crystal measuring 219.98 μm². In the higher magnification image, Figure 15(c) the crystals occupy 37.12% of the SEM image, totaling an area of 63.02 μm², with the largest crystal measuring 14.399 μm². These observations indicate a significant variation in crystal size and distribution across different magnifications and regions of the ZBLAN surface.

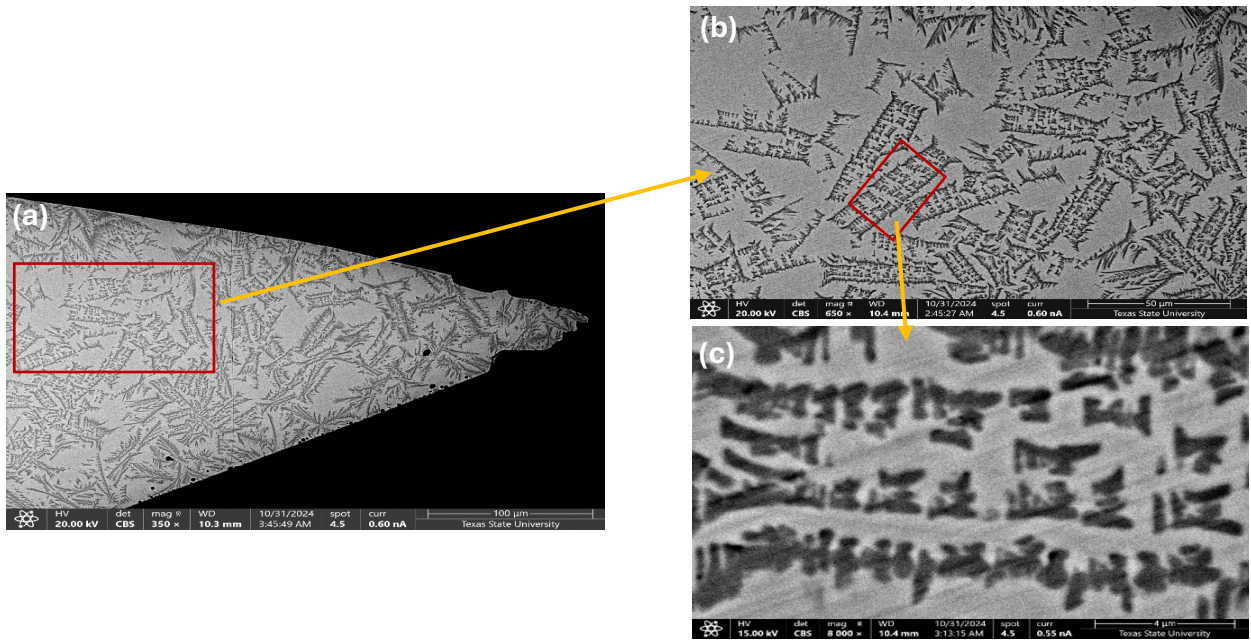

Figure 15: SEM images of ZBLAN treated at H1_360°C.

Figure 16 presents the SEM image of ZBLAN treated at a vibration of level H5 and temperature of 390°C. Crystals with different shapes were observed in various parts of the sample, as depicted in Figure 16(a) and Figure 16(b). In Figure 16(a), bow-tie-shaped crystals dominate, forming clusters that exhibit oval and rectangular shapes, highlighted in red and green, respectively. These crystals cover 9.45% of the SEM image area, amounting to a total of 1584.91 μm², with the largest crystal measuring 41.83 μm². Similarly, Figure 16(b) shows the presence of feather-shaped crystals along with bow-tie-shaped crystals in another region of the sample. These crystals cover 24.86% of the SEM image area, corresponding to 1182.58 μm², with the largest crystal having a size of 59.91 μm². These observations indicate a significant variation in crystal morphology and distribution across the ZBLAN surface under these conditions.

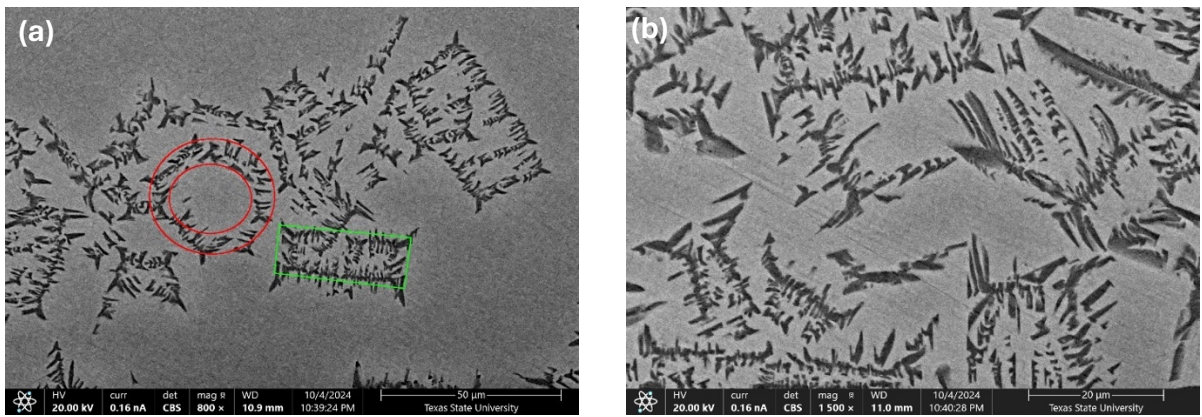

Figure 16: SEM images of ZBLAN treated at vibration frequency of level H5 and temperature of 390°C.

Overall, the SEM analysis demonstrates that vibration influences both the onset and morphology of crystallization in ZBLAN. Samples subjected to vibrational treatment exhibit earlier nucleation and distinct morphological transitions compared to those processed under purely thermal conditions. While certain trends suggest that higher vibration intensities promote greater nucleation density and morphological refinement, some variability was observed among samples, likely due to differences in local thermal contact and the short treatment duration. These observations, together with the reduction in crystallization onset temperature, indicate that vibration contributes directly to structural rearrangement and nucleation enhancement, although further quantitative analysis is required to isolate its precise role from concurrent thermal effects.

To complement the morphological findings obtained from SEM, energy-dispersive X-ray spectroscopy (EDS) was performed to examine the elemental distribution in ZBLAN samples subjected to different thermal and vibrational conditions. Two representative EDS results are presented below, one corresponding to an amorphous sample and the other to a crystalline sample. Comparison of these spectra allows evaluation of how the major constituent elements are distributed across the sample and how their spatial variation relates to the presence or absence of crystalline phases. This analysis thus serves to correlate structural transformations with compositional changes induced by vibration-assisted heat treatment.

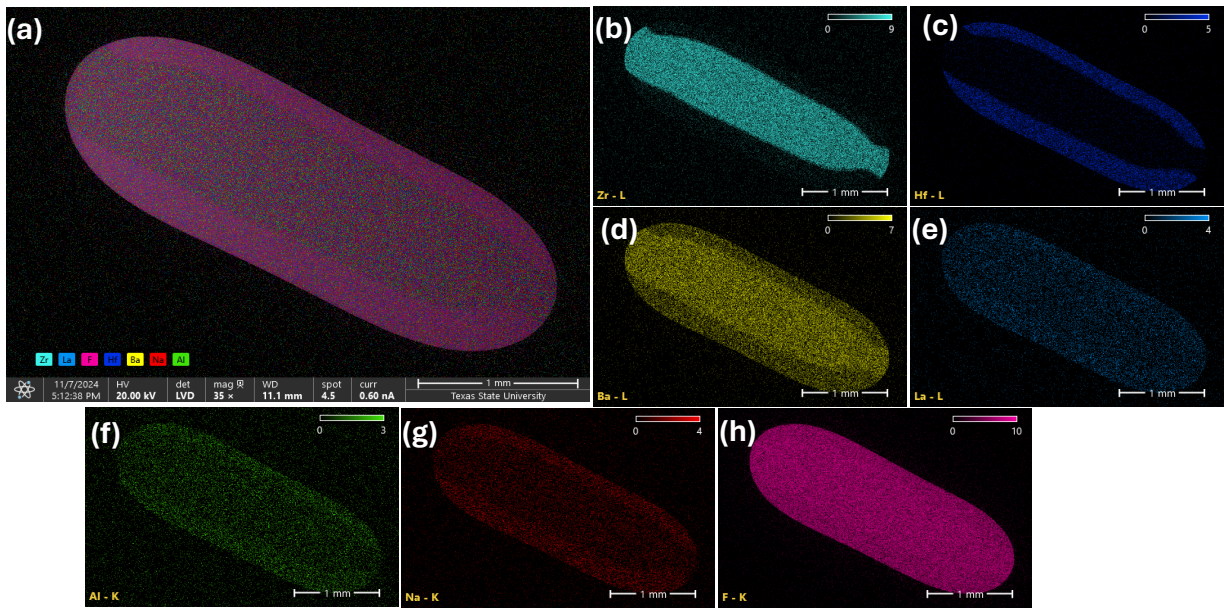

Figure 17: EDS count map of ZBLAN treated at vibration frequency of level H1 and temperature of 340°C.

Figure 17 presents the EDS results of ZBLAN treated at a vibration level of H1 and a temperature of 340°C, highlighting the elemental composition across different layers. The analysis identifies the presence of Zirconium (Zr), Hafnium (Hf), Barium (Ba), Lanthanum (La), Aluminum (Al), Sodium (Na), and Fluorine (F) in ZBLAN, as shown in Figure 17(b) through Figure 17(h) respectively.

As previously discussed, ZBLAN comprises two distinct layers: an inner core layer consisting of Zr and an outer cladding layer containing Hf. This layer differentiation is clearly depicted in Figure 17(b) and Figure 17(c), where Figure 17(b) reveals Zr presence in the inner core, and Figure 17(c) shows Hf concentrated in the outer cladding. Similarly, the concentration of Ba is higher in core while the concentration of La, Al and Na is higher in the cladding which is seen in Figure 17(d), Figure 17 (e), Figure 17(f) and Figure 17(g) respectively by the difference in the color. In contrast, fluorine (F) is distributed across both layers homogeneously, as indicated in the Figure 17(h).

Figure 18 presents a graphical representation of elemental composition versus weight percentage for the sample shown in Figure 17(a). The data reveal that F has the highest weight percentage, exceeding 25%, as it is a major constituent in both the cladding and core layers. Zr primarily present in the core, has a substantial weight percentage of approximately 24%. Ba and Hf, with Hf located in the cladding, each contribute around 20% by weight. In contrast, the remaining elements Na, La, and Al each exhibit weight percentages below 5%, reflecting their relatively lower concentration in ZBLAN.

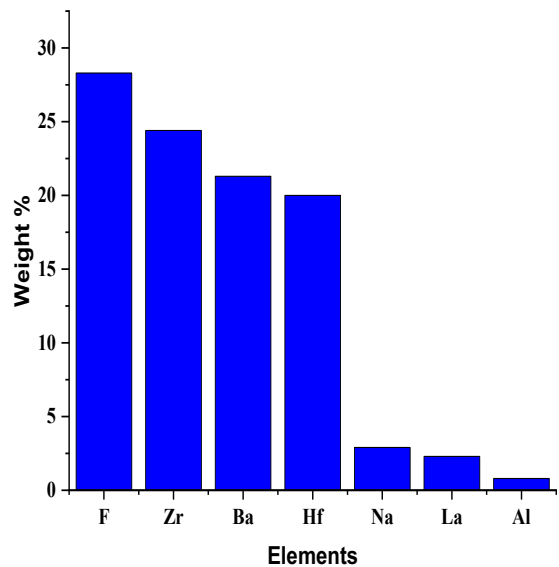

Figure 18: Elements Vs. Weight percentage of ZBLAN treated with vibration frequency of H1 and temperature of 340°C.

Figure 19 presents the EDS analysis of ZBLAN after treatment at temperature of 390°C, revealing a substantial increase in crystal formation on the surface. Figure 19(a) illustrates these crystals within the cladding layer of ZBLAN. As the cladding contains Hf exclusively, without Zr, no Zr is detected in this analysis. The EDS data show a high concentration of Hf within the crystals, while Ba and La are absent from the crystal surfaces, as shown in Figure 19(b), Figure 19(c), and Figure 19(d) respectively.

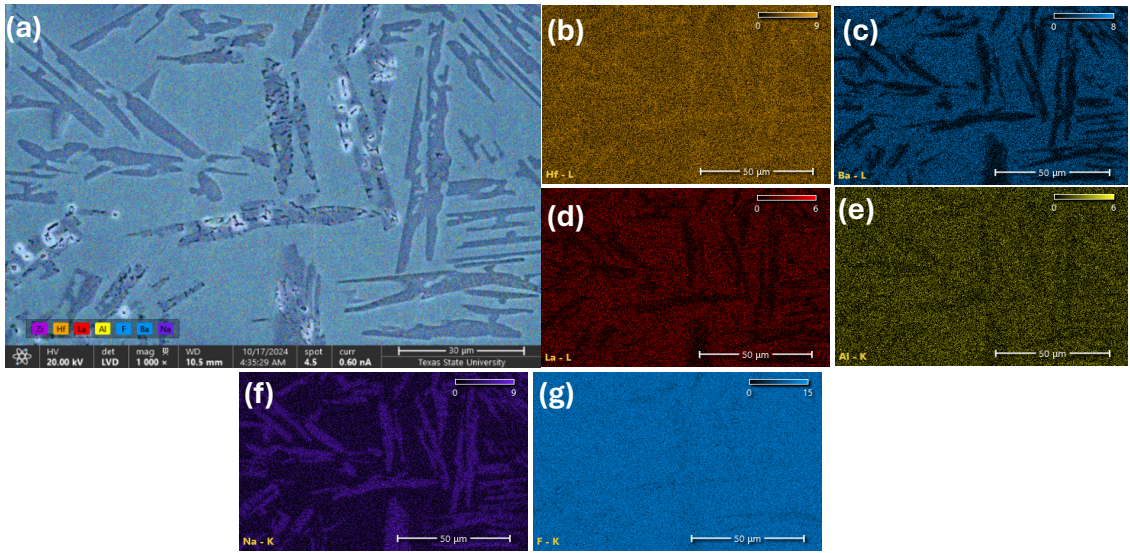

Figure 19: EDS count map of ZBLAN treated at temperature of 390°C.

Additionally, Al is lightly detected on the crystal surfaces, while Na appears in elevated concentrations, as depicted in Figure 19(e) and Figure 19(f) respectively. Likewise, F is uniformly distributed across the surface, as indicated in Figure 19(g). This analysis underscores the elemental distribution on the crystal surface within the cladding layer following thermal treatment.

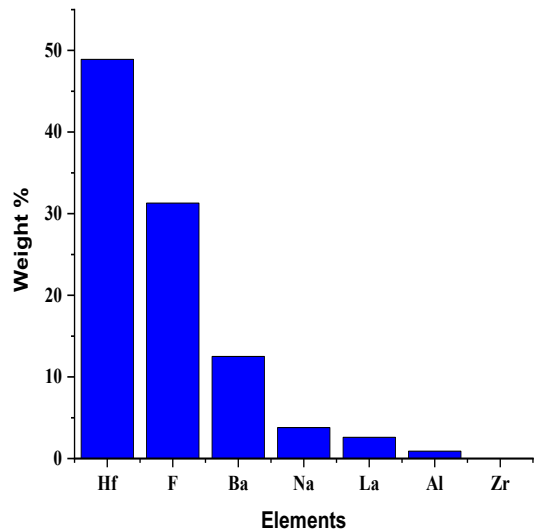

Figure 20: Elements Vs. Weight percentage of ZBLAN treated with temperature of 390°C.

The elemental weight percentage of the heat treated ZBLAN sample shown in Figure 19 is graphically represented in Figure 20. Here, Hf exhibits the highest concentration at nearly 50%, followed by F at approximately 30%. Ba constitutes around 10% of the composition, while the remaining elements each contributing less than 5% by weight include no detectable presence of Zr. This distribution highlights the dominance of Hf and F in the crystallized surface layer.

Overall, the EDS results corroborate the SEM observations by confirming distinct compositional differences between the amorphous and crystalline regions of ZBLAN. The enrichment of hafnium and fluorine in the crystallized surface layer, combined with the depletion of lighter elements such as sodium and aluminum, suggests preferential segregation of high atomic number constituents during crystal growth. These findings reinforce the conclusion that vibration-assisted heat treatment not only alters the microstructure but also influences elemental redistribution, promoting localized crystallization within the cladding region.

To further correlate the compositional and morphological observations obtained from SEM and EDS with surface topography, atomic force microscopy (AFM) analysis was conducted. This technique quantitatively characterizes the surface roughness and morphology of ZBLAN samples, enabling comparison between amorphous and crystallized regions and revealing how vibration-assisted heat treatment influences the microstructural texture of the glass.

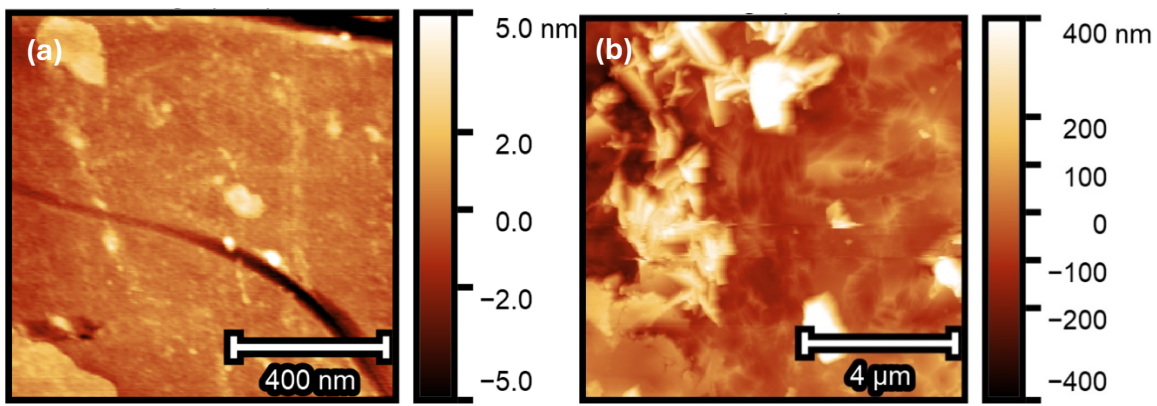

Figure 21: AFM image of ZBLAN (a) as received without any treatment, (b) after treating with a vibration Level of L3 and temperature of 390°C.

The AFM analysis reveals significant differences in surface morphology between untreated and treated ZBLAN samples. Figure 21(a) illustrates the AFM image of the as-received ZBLAN, showing a smooth surface with minimal crystal formation and a Root Mean Square (RMS) roughness of 1.168 nm. In contrast, Figure 21(b) displays the AFM image of the ZBLAN sample subjected to a vibration level of L3 and a temperature of 390°C, where notable crystal formation is evident. This treated sample exhibits a much higher surface roughness, calculated at 148.2 nm.

The comparison demonstrates that increasing crystallization results in a corresponding rise in surface roughness, confirming that vibration and temperature collectively enhance nucleation and crystal growth. This observation supports the conclusion that vibration-assisted thermal treatment not only influences internal crystallization but also significantly modifies the surface morphology of ZBLAN, producing a rougher and more heterogeneous topography consistent with increased crystallinity.

## Conclusion

A systematic investigation of ZBLAN under controlled thermal and vibrational conditions revealed the combined influence of temperature and mechanical excitation on its crystallization behavior. In temperature-only experiments, crystallization began between 330°C and 350°C and progressed steadily up to 400°C, establishing a consistent thermal baseline for comparison. The introduction of low-amplitude vibration produced results like those of purely thermal treatment, indicating that mild vibration does not significantly affect nucleation kinetics or heat-transfer behavior.

In contrast, high-vibration treatments led to an earlier onset of crystallization and more complex crystal morphologies, confirming that mechanical energy enhances atomic mobility and promotes nucleation. However, at higher vibration levels (H3, H4, and H5), several samples exhibited limited or delayed crystallization despite sufficient temperature exposure. This deviation is attributed to a jostling effect, where excessive vibration caused sample displacement within the silica ampoule, reducing thermal contact and hindering uniform heat transfer. Supporting video observations and preliminary COMSOL Multiphysics modeling validate this explanation and underscore the importance of maintaining stable thermal coupling during vibration-assisted processing.

Overall, this study demonstrates that controlled vibration can enhance crystallization in ZBLAN by accelerating atomic rearrangement and nucleation, while excessive vibration counteracts these benefits by disrupting heat transfer. These findings establish a framework for optimizing vibration-assisted crystallization in fluoride glasses, offering insights relevant to both terrestrial and microgravity manufacturing. Detailed COMSOL simulations and a redesigned experimental setup to further validate these conclusions will be presented in a forthcoming publication.

## Declaration of competing interest

The authors declare that they have no known competing financial interests or personal relationships that could have appeared to influence the work reported in this paper.

## Acknowledgement

This research was supported by the Universities Space Research Association (USRA). The authors gratefully acknowledge the Shared Research Operations (SRO) Center at Texas State University for providing access to SEM and related characterization facilities. The authors also thank Fiber Labs Inc. for supplying the ZBLAN preforms and AG Scientific for fabricating the silica ampoules used in this study.